\documentclass[prl,unsortedaddress,superscriptaddress,showpacs,twocolumn,floatfix]{revtex4}

\usepackage{graphicx}

\begin{document}

\title{Absolute Drell-Yan Dimuon Cross Sections in
       800 GeV/$c$ $pp$ and $pd$ Collisions}

\affiliation{Abilene Christian University, Abilene, TX 79699}
\affiliation{Argonne National Laboratory, Argonne, IL 60439}
\affiliation{Fermi National Accelerator Laboratory, Batavia, IL 60510}
\affiliation{Georgia State University, Atlanta, GA 30303}
\affiliation{Illinois Institute of Technology, Chicago, IL  60616}
\affiliation{Los Alamos National Laboratory, Los Alamos, NM 87545}
\affiliation{New Mexico State University, Las Cruces, NM 88003}
\affiliation{Oak Ridge National Laboratory, Oak Ridge, TN 37831}
\affiliation{Texas A\&M University, College Station, TX 77843}
\affiliation{Valparaiso University, Valparaiso, IN 46383}

\author{J.C.~Webb}
\altaffiliation[Present address: ]{Indiana University, Bloomington, IN 47405} 
\affiliation{New Mexico State University, Las Cruces, NM 88003}

\author{T.C.~Awes}
\affiliation{Oak Ridge National Laboratory, Oak Ridge, TN 37831}

\author{M.L.~Brooks}
\affiliation{Los Alamos National Laboratory, Los Alamos, NM 87545}

\author{C.N.~Brown}
\affiliation{Fermi National Accelerator Laboratory, Batavia, IL 60510}

\author{J.D.~Bush}
\affiliation{Abilene Christian University, Abilene, TX 79699}

\author{T.A.~Carey}
\affiliation{Los Alamos National Laboratory, Los Alamos, NM 87545}

\author{T.H.~Chang}
\altaffiliation[Present address: ]{University of Illinois, Urbana, IL 61801}
\affiliation{New Mexico State University, Las Cruces, NM 88003}

\author{W.E.~Cooper}
\affiliation{Fermi National Accelerator Laboratory, Batavia, IL 60510}

\author{C.A.~Gagliardi}
\affiliation{Texas A\&M University, College Station, TX 77843}

\author{G.T.~Garvey}
\affiliation{Los Alamos National Laboratory, Los Alamos, NM 87545}

\author{D.F.~Geesaman}
\affiliation{Argonne National Laboratory, Argonne, IL 60439}

\author{E.A.~Hawker}
\altaffiliation[Present address: ]{University of Cincinnati,
Cincinnati, OH 45221}
\affiliation{Texas A\&M University, College Station, TX 77843}
\affiliation{Los Alamos National Laboratory, Los Alamos, NM 87545}

\author{X.C.~He}
\affiliation{Georgia State University, Atlanta, GA 30303}

\author{L.D.~Isenhower}
\affiliation{Abilene Christian University, Abilene, TX 79699}

\author{D.M.~Kaplan}
\affiliation{Illinois Institute of Technology, Chicago, IL  60616}

\author{S.B.~Kaufman}
\affiliation{Argonne National Laboratory, Argonne, IL 60439}

\author{D.D.~Koetke}
\affiliation{Valparaiso University, Valparaiso, IN 46383}

\author{D.M.~Lee}
\affiliation{Los Alamos National Laboratory, Los Alamos, NM 87545}

\author{W.M.~Lee}
\altaffiliation[Present address: ]{Florida State University,
Tallahassee, FL 32306}
\affiliation{Georgia State University, Atlanta, GA 30303}

\author{M.J.~Leitch}
\affiliation{Los Alamos National Laboratory, Los Alamos, NM 87545}

\author{N.~Makins}
\altaffiliation[Present address: ]{University of Illinois, Urbana, IL 61801}
\affiliation{Argonne National Laboratory, Argonne, IL 60439}

\author{P.L.~McGaughey}
\affiliation{Los Alamos National Laboratory, Los Alamos, NM 87545}

\author{J.M.~Moss}
\affiliation{Los Alamos National Laboratory, Los Alamos, NM 87545}

\author{B.A.~Mueller}
\affiliation{Argonne National Laboratory, Argonne, IL 60439}

\author{P.M.~Nord}
\affiliation{Valparaiso University, Valparaiso, IN 46383}

\author{V.~Papavassiliou}
\affiliation{New Mexico State University, Las Cruces, NM 88003}

\author{B.K.~Park}
\affiliation{Los Alamos National Laboratory, Los Alamos, NM 87545}

\author{J.C.~Peng}
\altaffiliation[Present address: ]{University of Illinois, Urbana, IL 61801}
\affiliation{Los Alamos National Laboratory, Los Alamos, NM 87545}

\author{G.~Petitt}
\affiliation{Georgia State University, Atlanta, GA 30303}

\author{P.E.~Reimer}
\affiliation{Los Alamos National Laboratory, Los Alamos, NM 87545}
\affiliation{Argonne National Laboratory, Argonne, IL 60439}

\author{M.E.~Sadler}
\affiliation{Abilene Christian University, Abilene, TX 79699}

\author{W.E.~Sondheim}
\affiliation{Los Alamos National Laboratory, Los Alamos, NM 87545}

\author{P.W.~Stankus}
\affiliation{Oak Ridge National Laboratory, Oak Ridge, TN 37831}

\author{T.N.~Thompson}
\affiliation{Los Alamos National Laboratory, Los Alamos, NM 87545}

\author{R.S.~Towell}
\affiliation{Abilene Christian University, Abilene, TX 79699}
\affiliation{Los Alamos National Laboratory, Los Alamos, NM 87545}

\author{R.E.~Tribble}
\affiliation{Texas A\&M University, College Station, TX 77843}

\author{M.A.~Vasiliev}
\altaffiliation[On leave from ]{Kurchatov Institute, Moscow, Russia}
\affiliation{Texas A\&M University, College Station, TX 77843}

\author{J.L.~Willis}
\affiliation{Abilene Christian University, Abilene, TX 79699}

\author{D.K.~Wise}
\affiliation{Abilene Christian University, Abilene, TX 79699}

\author{G.R.~Young}
\affiliation{Oak Ridge National Laboratory, Oak Ridge, TN 37831}

\collaboration{FNAL E866/NuSea Collaboration}
\noaffiliation

\date{\today}

\begin{abstract}
The Fermilab E866/NuSea Collaboration 
has measured the Drell-Yan dimuon cross sections in 
800 GeV/$c$ $pp$ and $pd$ collisions.  This represents the first measurement 
of the Drell-Yan cross section in $pp$ collisions over a broad kinematic 
region and the most extensive study to date of the Drell-Yan cross section 
in $pd$ collisions.  The results indicate that recent global parton 
distribution fits provide a good description of the light antiquark sea in 
the nucleon over the Bjorken-$x$ range $0.03 \lesssim x < 0.15$, 
but overestimate the valence quark distributions as $x \to 1$.
\end{abstract}

\maketitle

Drell-Yan dimuon production provides valuable information about the partonic 
structure of hadrons that is complementary to deep-inelastic scattering (DIS) 
studies because it distinguishes between quarks and antiquarks.  
The Drell-Yan cross section for $p$A$\rightarrow\mu^+\mu^-X$ may be written
in terms of the parton distribution functions (PDFs) of the colliding
hadrons as
\begin{eqnarray}
  \label{eq:sigma}
   M^3 \frac{d^2\sigma}{dM\,dx_F} & = & \frac{8 \pi \alpha^2}{9} \frac{x_1 x_2}{x_1 + x_2}
   \\
   &\times & \sum_q e_q^2 \left[q_1(x_1) \bar{q}_2(x_2) + \bar{q}_1(x_1) q_2(x_2)\right] \nonumber
\end{eqnarray}
where the subscript 1\,(2) denotes the beam\,(target) hadron.  Measurement
of the invariant mass, $M^2=x_1x_2s$, and Feynman-$x$, 
$x_F=2p_L/\sqrt{s}=x_1-x_2$, of the muon pair
(where $p_L$ is the longitudinal momentum of the muon pair, and $\sqrt{s}$
the center-of-mass energy of the hadrons)
determines the momentum fraction $x_{1(2)}$ of the beam\,(target) parton.
Equation \ref{eq:sigma} is exact to leading order, but its general
features are also preserved by next-to-leading-order (NLO) calculations.

The Fermilab E866/NuSea Collaboration has measured the Drell-Yan cross 
sections in 800 GeV/$c$ $pp$ and $pd$ collisions.  This represents 
the first study of the Drell-Yan cross section in $pp$ collisions 
over a broad kinematic range and the most comprehensive measurement 
in $pd$ collisions.  
Previous publications
\cite{Hawk98,Tow01} have described the E866 measurement of the 
$\bar{d}/\bar{u}$ ratio in the proton as a function of $x$, based 
on the $x_2$ dependence of the Drell-Yan cross section ratio 
$\sigma^{pd}/2\sigma^{pp}$, and those results have been included in 
recent global PDF fits \cite{cteq5,cteq6,mrst98,mrst01,grv98}.

We present measurements of the $pp$ and $pd$
Drell-Yan dimuon cross sections for pair
mass in the ranges
$4.2<M<8.7$ GeV or $10.85<M<16.85$ GeV 
and $-0.05<x_F<0.8$.
At the large values of $x_F$
appropriate for much of the E866 data,
the first term
in the sum 
in Eq.\@ \ref{eq:sigma} dominates
and the cross sections are primarily sensitive to the valence 
distributions in the proton beam and the
antiquarks at small $x$
in the proton and deuteron targets.  
Thus, these measurements probe 
the partonic structure of the nucleon in two important kinematic regions.  
Recent global PDF fits use the precise HERA DIS cross sections to fix the 
magnitude of the light quark sea at small $x$ and the E605 $p$Cu cross 
sections \cite{e605} to fix the antiquark distributions for 
$0.14 \lesssim x \lesssim 0.3$.  In contrast, Ref.\@ \onlinecite{mrst02} noted that 
the magnitude of the antiquark distributions near 
$x \approx 0.04$ is relatively unconstrained by 
current data, 
and this limits the precision of 
calculations of the $W$ production cross section at the Tevatron.  There has 
also been considerable recent interest in understanding the valence quark 
distributions in the nucleon as $x \to 1$, which play a key role in searches 
for physics beyond the Standard Model, e.g., production of
additional vector bosons as predicted by left-right symmetric models.
To date, this region is constrained 
only by DIS cross sections, some of which suffer from limited statistics, 
while many others involve substantial ambiguities associated with corrections
for the nuclear targets used \cite{Yang99,Kuhl00}.  (See Ref.\@ \onlinecite{Zhang02} 
for a recent review.)  The present results provide new constraints on the 
magnitudes of the antiquark sea for $0.03 \lesssim x < 0.15$ and of the 
valence quarks for $x \to 1$.

E866 used a 3-dipole magnetic
spectrometer employed
previously in E605 \cite{e605}, E772 and E789, modified by the addition
of new detectors 
at the first tracking station.  
An 800 GeV/$c$ proton beam bombarded identical
target flasks containing
liquid hydrogen, liquid deuterium and vacuum that were alternated
every few minutes.  After passing
through the target, the beam was intercepted by a copper
beam dump that was
followed by a thick hadron absorber,
ensuring that only muons traversed the spectrometer's detectors.
Drell-Yan events were recorded using three
different
spectrometer magnet settings, chosen to focus low-, intermediate- and high-mass
muon pairs and provide acceptance from below the
$J/\psi$ to above 15 GeV.  A detailed description may be found in Ref.\@ \onlinecite{Tow01}.

The present analysis used 55,000 $pp$ and 121,000 $pd$ Drell-Yan events,
approximately half the statistics of the E866 
$\bar{d}/\bar{u}$ study \cite{Tow01}.  The
$\bar{d}/\bar{u}$ analysis was optimized 
to achieve minimum relative uncertainties in $\sigma^{pd}/2\sigma^{pp}$, 
whereas the present analysis was optimized for absolute measurements
of the cross sections.
Therefore, more stringent fiducial cuts were adopted in
the present analysis,  
eliminating events for which the absolute acceptance of the spectrometer 
could not be reliably determined.
This minimized systematic uncertainties at the cost of statistical
precision, especially for $x_2 \gtrsim 0.15$.
In contrast, 7\% of the events accepted in this analysis came from
data sets that were excluded previously \cite{Tow01,jason}.
Several improvements were made 
to the event reconstruction, notably involving the treatments of energy loss 
and multiple scattering in the absorber,
that led to a more precise reconstruction of the dimuon 
kinematics and a better match between the real and simulated events.

The spectrometer acceptance was calculated separately for each data set as a 
function of mass, $x_F$, and transverse momentum ($p_T$) using a detailed 
Monte Carlo simulation of the spectrometer.  The virtual photon azimuthal 
production and decay angles were assumed to be distributed isotropically, and 
the polar decay angles were assumed to be distributed according to 
$1 + \cos^2 \theta_d$, consistent with theoretical expectations and previous 
Drell-Yan angular distribution studies \cite{McGau99,Chang99,Bro01}.  
Drell-Yan Monte Carlo events were thrown with realistic kinematic 
distributions in mass, $x_F$, and $p_T$, then reweighted to provide a 
precise match to the observed experimental distributions. 
Triply-differential 
cross sections, $d^3\sigma/dM dx_F dp_T$, were calculated for each kinematic 
bin, then integrated over $p_T$ to obtain the invariant cross sections,
$M^3 d^2\sigma/dM dx_F$.  The transverse momentum acceptance of the 
spectrometer extended to
$p_T \approx 5-7$ GeV/$c$,
but it was nonetheless necessary to extrapolate the integration to large $p_T$ where the Monte Carlo showed zero acceptance.  The extrapolation contributed well under 1\%
in most cases, and was never more than 5\%.

A detailed study was performed of the 
point-to-point systematic uncertainties in the measured cross sections.  
The two dominant contributions were the statistical uncertainty in the Monte 
Carlo event samples that were used for the acceptance calculations and the 
absolute field strength of the spectrometer magnets.  Smaller contributions 
came from the uncertainties in the hodoscope, drift chamber, and trigger 
efficiencies, the composition and density of the targets, and the 
extrapolations to large $p_T$.  Finally, some large-$x_F$ events contained a 
muon that passed very close to the edge of the beam dump, increasing the 
uncertainty in the energy loss and multiple scattering corrections.  An 
additional $\pm$\,5\% systematic uncertainty was added to the final results 
for the affected kinematic bins.
The total point-to-point systematic uncertainties within any
($M$,$x_F$) bin were strongly correlated between the $pp$ and $pd$ cross
sections.
In addition to the point-to-point systematic uncertainties, there was a 
$\pm$\,6.5\% overall normalization uncertainty, associated with the 
calibration of the beam intensity.
See Ref.\@ \onlinecite{jason} for additional details.

\begin{figure*}[tb]
  \includegraphics*[width=\linewidth]{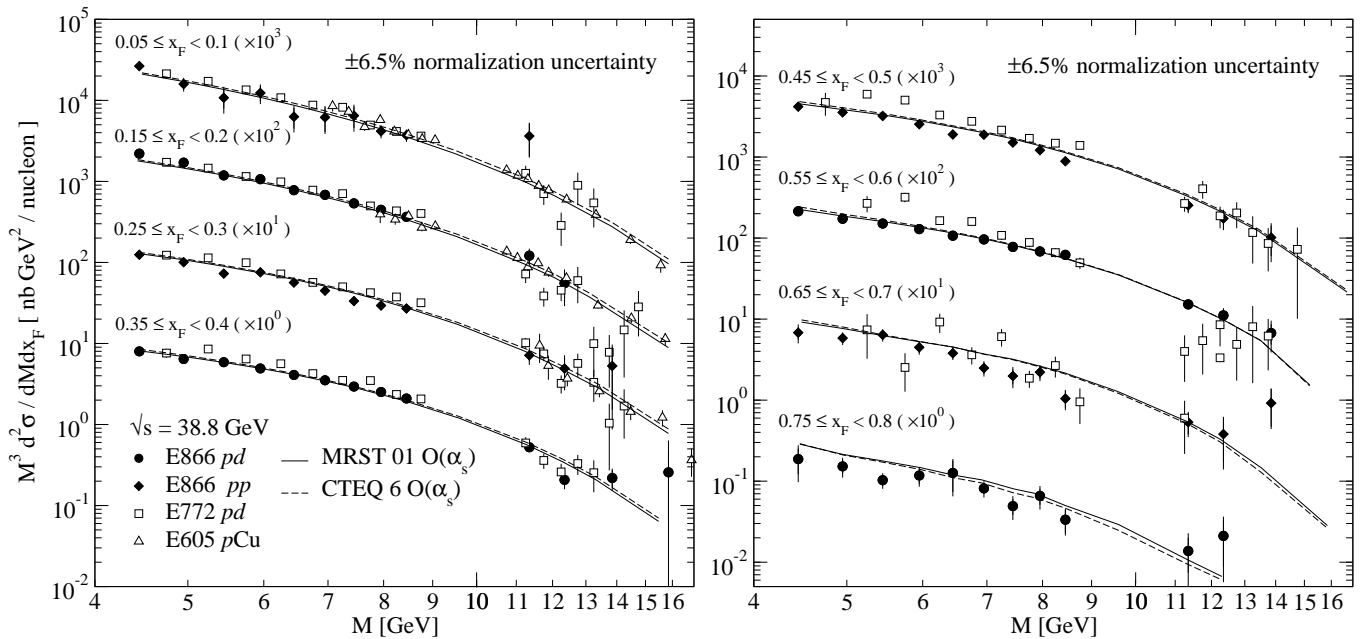}
  \caption{\label{fig:xfbins}
     FNAL E866 Drell-Yan cross sections per nucleon
     for selected $x_F$ bins.
     The E866 $pd$ (solid circles) and $pp$ (solid diamonds)
     cross sections are shown in
     alternate decades,
     compared with previous $p$Cu results from E605 \protect\cite{e605}
     (open triangles) and $pd$ results from E772 \protect\cite{e772} 
     (open squares).
     NLO cross section calculations based on the  CTEQ6 \protect\cite{cteq6} 
     (dashed curves) and MRST2001 \protect\cite{mrst01} (solid curves) PDF fits
     are also shown.  The error bars on the E605 and E772 data points are statistical 
     only.  Those on the E866 data points are the sum in quadrature of the statistical
     and point-to-point
     systematic uncertainties.  An additional $\pm$\,6.5\% 
     normalization uncertainty is common to all E866 data points.
     }
\end{figure*}

Figure \ref{fig:xfbins} shows the Drell-Yan $pp$ and $pd$ invariant cross 
sections per nucleon
for selected $x_F$ bins.  
The results agree with previous 800 GeV/$c$
Drell-Yan cross section measurements in $p$Cu collisions by E605 \cite{e605}.  
They also agree with previous measurements in $pd$ collisions by E772 
\cite{e772} for $x_F<0.3$.  At larger $x_F$ and small $M$, the E772 cross 
sections are systematically larger than the present results.  E772 quoted a 
larger point-to-point
systematic uncertainty in this kinematic region, but the results of the 
two experiments differ by more than the combined systematic uncertainties 
would predict.  The largest differences are in the region $0.4<x_F<0.6$ and 
$4.2<M<7$ GeV, where inconsistencies between the E772 results and expectations 
from global PDF fits were noted previously \cite{mrst98}.  This region was 
studied during E866 with very different acceptances by 
the low- and high-mass spectrometer settings, and the results
are consistent.

\begin{table}[tb]
\caption{\label{table:kprime}
$K^{\prime}$-factors obtained with various PDFs, where
$K^{\prime} = \sigma^{\rm exp}/\sigma^{\rm NLO}$.  The $\chi^2$
values include the statistical and point-to-point
systematic uncertainties in the experimental
cross sections, but not the $\pm\,6.5$\% global normalization uncertainty.
The $K^{\prime}$ fits have 183 and 190 degrees of freedom ($dof$) for the
$pp$ and $pd$ reactions, respectively.}
\begin{ruledtabular}
\begin{tabular}{lcccccc}
PDF       & \, & $K^{\prime}_{pp}$ & $\chi^2/dof$ & \, & $K^{\prime}_{pd}$ & $\chi^2/dof$ \\
\hline
CTEQ5 \protect\cite{cteq5}     & \, & 0.976 & 1.42 & \, & 0.963 & 2.51 \\
CTEQ6 \protect\cite{cteq6}     & \, & 1.016 & 1.39 & \, & 1.001 & 2.56 \\
MRST98 \protect\cite{mrst98}   & \, & 0.973 & 1.38 & \, & 0.960 & 2.37 \\
MRST2001 \protect\cite{mrst01} & \, & 0.980 & 1.45 & \, & 0.966 & 2.44 \\
GRV98 \protect\cite{grv98}     & \, & 0.811 & 2.04 & \, & 0.808 & 4.15 \\
\end{tabular}
\end{ruledtabular}
\end{table}

Figure \ref{fig:xfbins} also shows the results of next-to-leading-order 
calculations of the Drell-Yan cross sections based on the CTEQ6 \cite{cteq6} 
and MRST2001 \cite{mrst01} global PDF fits.  The agreement with the global 
fits is very good over the entire kinematic region.  This agreement may be 
quantified by computing a $K^{\prime}$-factor, which we define to be the 
ratio of the experimental cross section to a NLO prediction.  Table 
\ref{table:kprime} shows $K^{\prime}$-factors for several recent global PDF 
fits.  With the exception of GRV98, all of the recent PDF fits predict the 
absolute magnitude of the Drell-Yan cross sections to within the $\pm$\,6.5\% 
normalization uncertainty.

\begin{figure}[tb]
 \includegraphics*[width=8.5cm]{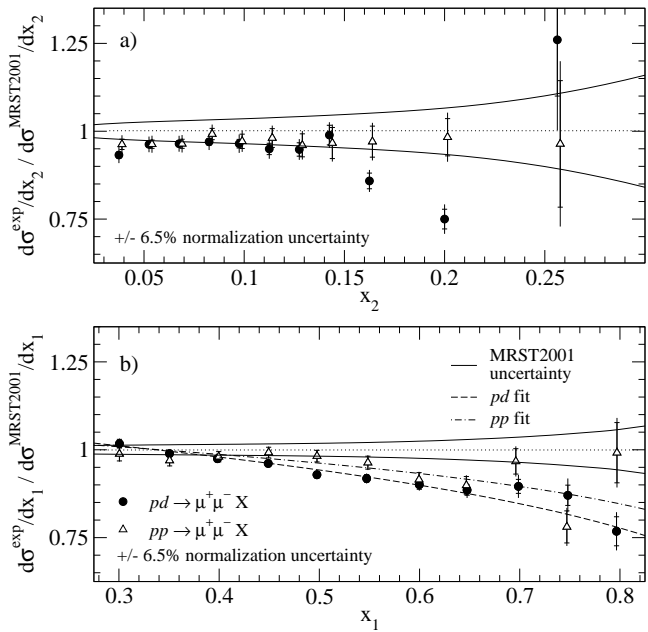}
  \caption{\label{fig:pdfcompare}
  Ratios of the measured Drell-Yan $pp$ (open triangles) and $pd$ 
  (solid circles) cross sections to NLO calculations based on the
  MRST2001 \protect\cite{mrst01} PDF fit plotted \textit{vs}.\@ a) $x_2$
  and b) $x_1$, averaged over the other variable.  The inner
  error bars represent the statistical uncertainty in the ratio, the
  outer error bar the sum in quadrature of the statistical and 
  point-to-point systematic uncertainties.  Solid lines represent
  the 
  experimental uncertainty ranges \protect\cite{mrst02}
  on a) $\bar{u}+\bar{d}$ and b) $4u+d$ in the
  MRST2001 PDF fit.
  Dashed and dot-dashed lines are the $pd$ and $pp$
  fits described in the text.}
\end{figure}

While the overall normalization is well reproduced, there are systematic 
deviations between the measurements and the predictions that are reflected 
in the large $\chi^2$ values.  To elucidate these deviations, it is useful 
to examine the experimental cross sections separately as functions of $x_1$ 
and $x_2$.  Most of the events have $x_1 \gg x_2$, which 
implies that the $x_1$ dependence is primarily sensitive to the valence quarks 
in the proton beam and the $x_2$ dependence to the antiquarks in the 
target.  
Figure \ref{fig:pdfcompare} shows the ratios of the experimental 
cross sections to NLO calculations using the MRST2001 global PDF 
fit, 
separately as a function of $x_2$ and $x_1$, averaged over the
other momentum fraction.  The uncertainties in the NLO calculations from the
PDF fit are also shown.
After accounting for the normalization uncertainty,
the MRST2001 partons provide a very good description of the $x_2$ dependence
of the $pp$ cross sections over the full range, and a good description
of the $pd$ cross sections for $x_2<0.15$.  The CTEQ6 fits describe the
$x_2$ dependence equally well.
This indicates that the current PDFs interpolate the 
$\bar{u}$ and $\bar{d}$ distributions successfully between the HERA 
measurements at small $x$ and the E605 measurements for $x>0.14$,
but these new results provide tighter constraints on the magnitude 
of the antiquark distributions for 
$0.03 \lesssim x < 0.15$ than have existed to date \cite{mrst02}.

Figure \ref{fig:pdfcompare}b shows the ratios of the cross sections to
NLO calculations using the MRST2001 PDF fits as a function of $x_1$.
The qualitative behavior of the CTEQ6 distributions is quite similar: 
both PDF fits overestimate the
valence quarks by $\approx$\,15--20\% at large $x_1$,
and the discrepancy between the data and current PDFs appears to
be larger for the $pd$ cross sections than for the $pp$ cross sections.  
The acceptance of these data includes an anti-correlation between
$x_1$ and $x_2$ \cite{Tow01}, so most of the events at
large $x_1$ also have $x_2<0.1$.  The NLO calculations
describe the average $x_2$ dependence of the cross sections
quite well for $x_2<0.1$, so the discrepancy in
Fig.\@ \ref{fig:pdfcompare}b is isolated to the
subset of the low-$x_2$ events that also have large $x_1$.  
Further comparisons between the measured cross sections and the NLO
calculations, in which the $x_1$ and $x_2$ dependences are examined for
limited ranges of the other variable \cite{jason}, reinforce this
conclusion.
Table \ref{table:fit} shows the results of fits 
to the $x_1$ dependence
of the $K^\prime$-factors with a phenomenological form, 
$K^\prime(x_1) = \alpha \left( 1 - x_1 \right)^\beta$, for both PDF sets.

\begin{table}[tb]
  \caption{\label{table:fit}Fits of data/theory to the phenomenological
   form $K^\prime(x_1) = \alpha \left( 1 - x_1 \right)^\beta$.  
   Uncertainties on the fits include only the statistical uncertainties in the 
   data.}
  \begin{ruledtabular}
  \begin{tabular}{lcccc}
            & \multicolumn{2}{c}{MRST2001} & \multicolumn{2}{c}{CTEQ6} \\
            & $pp$           & $pd$           & $pp$       & $pd$      \\
            \hline
   $\alpha$ & $1.06\pm 0.02$ & $1.09\pm 0.01$ & $1.11\pm 0.02$ & $1.15\pm 0.01$ \\
   $\beta$  & $0.14\pm 0.03$ & $0.21\pm 0.02$ & $0.15\pm 0.03$ & $0.23\pm 0.02$ \\
  \end{tabular}
  \end{ruledtabular}
\end{table}

Unlike large-$x$ DIS off deuterium targets, 
these data are in a kinematic region where nuclear-dependence corrections
are known to be small \cite{Alde90}.  
The $pp$ and $pd$ cross sections at large $x_1$ constrain slightly 
different linear combinations of $u(x)$ and $d(x)$,
with greater sensitivity to $u$ quarks.  The results imply that the 
$u$ quark distributions in CTEQ6 and MRST2001 are overestimated as $x \to 1$.  
Recent fits to H1 charged- and neutral-current data have also indicated that 
the global PDF fits appear to overestimate the $u$ quark distribution near 
$x\approx 0.65$, albeit with limited statistics \cite{Zhang02}.  The Drell-Yan 
cross sections may also point to problems with the $d/u$ ratio as $x \to 1$ 
\cite{Yang99,Kuhl00}.  However, determination of the $d/u$ ratio at large $x$ from 
the present data will require a fit to the full two-dimensional invariant 
cross sections together with the rest of the current world data.

Figure \ref{fig:pdfcompare} shows that the present results fall well outside the experimental uncertainty \cite{mrst02} for the valence quarks in the MRST2001 PDF fit.  A similar comparison shows that the present results track the lower limit on the experimental uncertainty \cite{cteq6} in the CTEQ6 PDF fit.
These discrepancies
reflect additional
uncertainties in the PDF fits
\cite{cteq6,mrst02}, and they imply that future PDF fits will see a substantial correction to the $u$ and $d$ quark distributions at large $x$.
The reduced valence quark distributions at large $x$ implied by these results will also lead to predictions of smaller cross sections for some signatures for new physics that are being sought at the Tevatron.

We thank the Fermilab Particle Physics, Beams, and
Computing Divisions for their assistance in performing this experiment.  
We also thank W.K.\@ Tung of the CTEQ Collaboration for providing the code to 
calculate the next-to-leading-order Drell-Yan cross sections.
This work was supported in part by the
U.S.\@ Department of Energy.


\begin{thebibliography}{99}
\bibitem{Hawk98} E.A. Hawker \textit{et al}.\@ (FNAL E866/NuSea Collaboration),
Phys. Rev. Lett. \textbf{80}, 3715 (1998).
\bibitem{Tow01} R.S. Towell \textit{et al}.\@ (FNAL E866/NuSea Collaboration),
Phys. Rev. D \textbf{64}, 052002 (2001).
\bibitem{cteq5} H.L. Lai \textit{et al}., Eur. Phys. J. C \textbf{12}, 375 (2000).
\bibitem{cteq6} J. Pumplin \textit{et al}., JHEP \textbf{0207}, 012 (2002).
\bibitem{mrst98} A.D. Martin, R.G. Roberts, W.J. Stirling, and R.S. Thorne, Eur. Phys. J. C \textbf{4}, 463 (1998).
\bibitem{mrst01} A.D. Martin, R.G. Roberts, W.J. Stirling, and R.S. Thorne, Eur. Phys. J. C \textbf{23}, 73 (2002).
\bibitem{grv98} M. Gluck, E. Reya, and A. Vogt, Eur. Phys. J. C \textbf{5}, 461 (1998).
\bibitem{e605} G. Moreno \textit{et al}.,
Phys. Rev. D \textbf{43}, 2815 (1991).
\bibitem{mrst02} A.D. Martin, R.G. Roberts, W.J. Stirling, and R.S. Thorne, hep-ph/0211080.
\bibitem{Yang99} U.K. Yang and A. Bodek, Phys. Rev. Lett. \textbf{82}, 2467 (1999).
\bibitem{Kuhl00} S. Kuhlmann \textit{et al}., Phys. Lett. \textbf{B476}, 291 (2000).
\bibitem{Zhang02} Z. Zhang (H1 and ZEUS Collaborations), J. Phys. G \textbf{28}, 767 (2002).
\bibitem{jason} J.C. Webb, Ph.D.\@ thesis, New Mexico State University (2002);
hep-ex/0301031.
\bibitem{McGau99} P.L. McGaughey, J.M. Moss, and J.C. Peng, Ann. Rev. Nucl. Part. Sci. \textbf{49}, 217 (1999).
\bibitem{Chang99} T.H. Chang, Ph.D.\@ thesis, New Mexico State University (1999);
hep-ex/0012034.
\bibitem{Bro01} C.N. Brown \textit{et al}.\@ (FNAL E866/NuSea Collaboration),
Phys. Rev. Lett. \textbf{86}, 2529 (2001).
\bibitem{e772} P.L. McGaughey \textit{et al}.,
Phys. Rev. D \textbf{50}, 3038 (1994); \textbf{60}, 119903 (1999).
\bibitem{Alde90} D.M. Alde \textit{et al}.,
Phys. Rev. Lett. \textbf{64}, 2479 (1990).
\end{thebibliography}
\end{document}